\documentclass[conference,lettersize]{IEEEtran}
\IEEEoverridecommandlockouts
\usepackage{cite}
\usepackage{comment}
\usepackage{amsmath,amssymb,amsfonts}
\usepackage{algorithmic}
\usepackage{graphicx}
\usepackage{float}
\usepackage{textcomp}
\usepackage{xcolor}
\usepackage{subfigure}
 
\usepackage{url}
\usepackage{algorithm}
\usepackage{algorithmic}

\usepackage[top=.6in,left=.625in,right=.625in,bottom=.9in]{geometry}

\def\BibTeX{{\rm B\kern-.05em{\sc i\kern-.025em b}\kern-.08em
    T\kern-.1667em\lower.7ex\hbox{E}\kern-.125emX}}

\usepackage{fancyhdr}

\begin{document}




\title{\fontsize{22}{24}\selectfont  Resource~\hspace{2pt}Allocation~\hspace{2pt}for~\hspace{2pt}Mobile~\hspace{2pt}Metaverse~\hspace{2pt}with~\hspace{2pt}the Internet~\hspace{2pt}of~\hspace{2pt}Vehicles~\hspace{2pt}over~\hspace{2pt}6G~\hspace{2pt}Wireless~\hspace{2pt}Communications: \\A~\hspace{2pt}Deep~\hspace{2pt}Reinforcement~\hspace{2pt}Learning~\hspace{2pt}Approach}

\author{\IEEEauthorblockN{Terence Jie Chua}
\IEEEauthorblockA{\textit{Graduate College}\\\textit{Nanyang Technological University}\\Singapore\\
terence001@e.ntu.edu.sg }
\and

\IEEEauthorblockN{Wenhan Yu}
\IEEEauthorblockA{\textit{Graduate College}\\\textit{Nanyang Technological University}\\Singapore\\
wenhan002@e.ntu.edu.sg }
\and

\IEEEauthorblockN{Jun Zhao}
\IEEEauthorblockA{\textit{School of Computer Science \& Engineering}\\ \textit{Nanyang Technological University}\\Singapore\\
junzhao@ntu.edu.sg }
}

\maketitle

 \thispagestyle{fancy}
\pagestyle{fancy}
\lhead{This paper appears in the Proceedings of 8th IEEE World Forum on the Internet of Things (\textbf{WFIoT}) 2022.\\ Please feel free to contact us for questions or remarks.} 

\cfoot{~\\[-30pt]\thepage}
  




\begin{abstract}
Improving the interactivity and interconnectivity between people is one of the highlights of the Metaverse. The Metaverse relies on a core approach, digital twinning, which is a means to replicate physical world objects, people, actions and scenes onto the virtual world. Being able to access scenes and information associated with the physical world, in the Metaverse in real-time and under mobility, is essential in developing a highly accessible, interactive and interconnective experience for all users. This development allows users from other locations to access high-quality real-world and up-to-date information about events happening in another location, and socialize with others hyper-interactively. Nevertheless, receiving continual, smooth updates generated by others from the Metaverse is a challenging task due to the large data size of the virtual world graphics and the need for low latency transmission. With the development of Mobile Augmented Reality (MAR), users can interact via the Metaverse in a highly interactive manner, even under mobility. Hence in our work, we considered an environment with users in moving Internet of Vehicles (IoV), downloading real-time virtual world updates from Metaverse Service Provider Cell Stations (MSPCSs) via wireless communications. We design an environment with multiple cell stations, where there will be a handover of users' virtual world graphic download tasks between cell stations. As transmission latency is the primary concern in receiving virtual world updates under mobility, our work aims to allocate system resources to minimize the total time taken for users in vehicles to download their virtual world scenes from the cell stations. We utilize a deep reinforcement learning approach and evaluate the performance of the algorithms under different environmental configurations. Our work provides a use case of the Metaverse over AI-enabled 6G wireless communications.
\end{abstract}







\section{Introduction}

One of the foundations for developing the Metaverse is digital twinning \cite{lee2021all}, in which real-world objects, people, actions and events are mapped to and replicated in the virtual world. One of the perks of having an integrated physical and virtual world is that it opens doors to achieving greater accessibility, interconnectivity and interactivity between people and places, allowing users to access physical world and real-time information changes, and connect with others on a much more personal level through augmented reality (AR) \cite{lee2021all}.

\textbf{Motivation.} In the age of mobile edge devices, interacting with others and accessing real-world, real-time information via mobile devices under mobility is proliferated. However, obtaining these virtual world AR real-time information updates and interactions under mobility over wireless communications can be challenging, as this information is downloaded in a graphical format that can be of large data size. With the continuous stream of data downloaded from the Metaverse Service Provider Cell Station (MSPCS) and the continuous movement of real-world users, there is concern about the virtual world scenes download latency. 
\begin{figure}[t]
\centering
\includegraphics[width=1\linewidth]{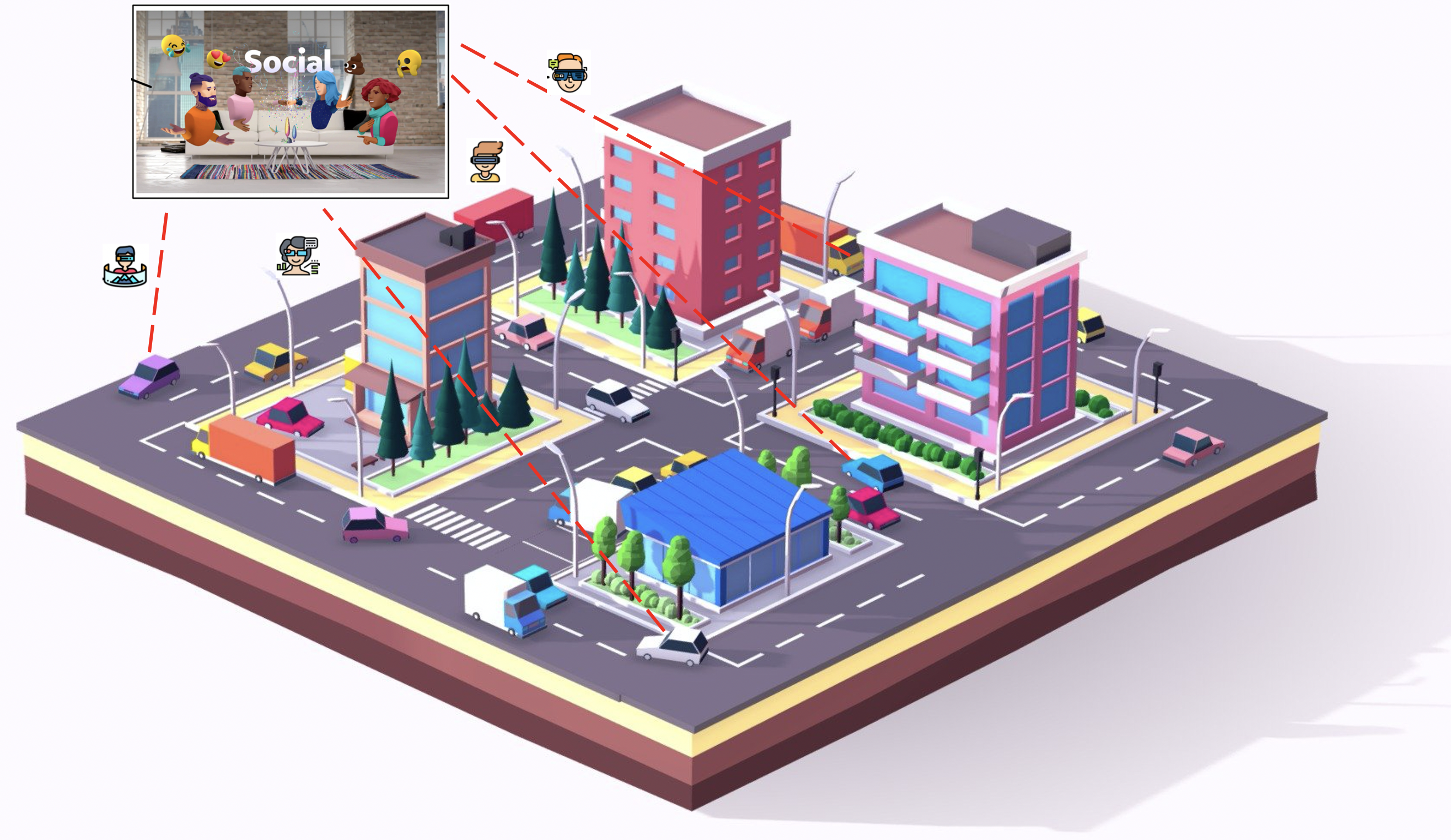}
\caption{Mobile users travelling in the Internet of Vehicles (IoV) downloading virtual world scenes under mobility.}
\label{fig:scene}
\vspace{-0.4cm}
\end{figure}

\textbf{Related work.} 
Since the Metaverse is still relatively new, limited studies consider the user device-Metaverse edge server communication and computation framework. Han~\textit{et~al.}~\cite{han2022dynamic} proposed a resource allocation framework for the Internet of Things (IoT) to facilitate the synchronization of the Metaverse with the physical world. The above-mentioned work mainly utilized game-theoretic approaches to tackle their defined problem.
Ng \textit{et al.}~\cite{ng2022unified} proposed a framework that uses stochastic optimization based resource allocation to obtain the minimum cost of a Metaverse   service provider in the education sector context. Xu~\textit{et~al.}~\cite{xu2022wireless} introduced an incentive mechanism based on machine learning for VR in the Metaverse, using auction theory to obtain the optimal pricing and allocation, and deep reinforcement learning to accelerate the auction process.
Li~\textit{et~al.}~\cite{li2022internet} surveyed how the IoT can be combined with the Metaverse.
Yang~\textit{et~al.}~\cite{yang2022metafi} proposed an IoT-enabled pose estimation scheme for the Metaverse.

There are several works~\cite{chen2017resource,wang2021meta,liu2018edge,wang2020user} which have studied resource management or optimization problems concerning virtual, augmented or extended reality over wireless networks. Chen~\textit{et~al.}~\cite{chen2017resource} considered a resource allocation problem and devised a distributed machine learning algorithm to tackle it. Wang \textit{et al.}~\cite{wang2021meta} 
introduced an indoor virtual reality scenario in which they utilized reinforcement learning to improve convergence speed and the sum of successful transmission probabilities by optimizing VLC access points (VAP) and user-base station association. Liu \textit{et al.}~\cite{liu2018edge} proposed an algorithm to tackle the trade-off between network latency and video object detection accuracy for mobile augmented reality systems. Wang \textit{et al.}~\cite{wang2020user}  aimed to minimize the energy consumption of users using mobile augmented reality systems by optimizing the MAR configuration as well as the radio resource allocation. In \cite{chen2018virtual,guo2020adaptive}, the authors studied optimization problems which consider the Quality of Experience (QoE) of users, with Chen \textit{et al.}~\cite{chen2018virtual} using a game theoretic approach, and Guo~\textit{et~al.}~\cite{guo2020adaptive} proposing both game theoretic and reinforcement learning approaches to tackle their defined problem.

While many works \cite{liu2018edge,tong2016hierarchical,tan2017online,guo2020adaptive,wang2021meta} have considered variants of user to edge server allocation optimization problems, these works did not consider the application of deep reinforcement learning in the context of bettering the interconnectivity, accessibility, socialization through interactive AR Metaverse. 


Our use of artificial intelligence (AI) for the Metaverse over wireless communications is a pioneering use case of the sixth-generation
(6G) wireless communications, which are being actively developed worldwide. Interested readers can refer to~\cite{zhao2021survey,yang20196g,akyildiz20206g} for more discussions of 6G. The application of AI, as in our paper, can improve the system performance and make automated decisions for 6G to enable high-end applications like the Metaverse which requires low latency and high data rate.

\textbf{Our approach.} 
We propose a simulated environment in which user equipments (UEs) in the Internet of Vehicles (IoV) are traversing the city and are requesting real-time virtual world updates and graphics from the Metaverse Service Provider Cell Station (MSPCS), as illustrated in Fig.~\ref{fig:scene}. The vehicles in our simulated environment are constantly moving. The distance between UEs in cars and the MSPCSs are constantly changing, resulting in changing channel gain between the UEs and the MSPCSs, and consequently, varying data transfer rates. The main goal of our work is to minimize the total time users in vehicles take to download a set of virtual world AR graphical scenes from the MSPCS wireless communications. As such, we introduced a reinforcement learning-based UE-MSPCS orchestrator which aims to minimize total data download delay by optimizing the UE-MSPCS allocation. For a given set of virtual world AR graphical scenes to be downloaded, the size of graphical data remaining to be downloaded keeps changing with each step that the users take (as the users are downloading at each time step). This optimization problem is not suitable to solve via standard optimization procedures due to the sequential nature of the problem.

\textbf{Contributions.} Our contributions are as follows:
\begin{itemize}
\item \emph{\textbf{Problem formulation of multi-user, multi-MSPCS virtual world scene downloading under mobility:}} We present a novel Metaverse socialization framework under user mobility, and introduce a deep reinforcement learning-based user-to-MSPCS orchestrator which aims to minimize the latency of data transfer between MSPCS and users travelling within the Internet of Vehicles (IoV), as shown in Fig~\ref{fig:model}. Our work introduces a solution to carrying out real-time on-the-go virtual environment interactions and updates in the real-world, taking into consideration geographical locations of users in vehicles and MSPCS.

\item \emph{\textbf{Reinforcement learning in a Metaverse scenario:}} We champion the application of AI, in particular, deep reinforcement learning, to resource allocation for the Metaverse over wireless communications.

\item \emph{\textbf{Inter-cell and Intra-cell interference:}} In our work, we consider both intra-cell (interference as a result of a single cell transmitting signals to different users) and inter-cell (interference as a result of multiple cells transmitting signals to different users).

\item \emph{\textbf{Analyses of varying reinforcement learning algorithms and environment setting:}} We employ varying state-of-the-art reinforcement learning algorithms such as PPO, DQN, Dueling DQN and A2C, and provide an in-depth comparisons and analyses.

\item \emph{\textbf{A use case for the Metaverse over AI-enabled 6G wireless communications:}} 6G wireless communications have recently attracted much attention. We believe AI will be a crucial building block for 6G wireless communications. Our approach of using reinforcement learning for Metaverse AR resource allocation over wireless communications can   be seen as a use case of the Metaverse over 6G.
 
\end{itemize}

The rest of the paper is organized as follows. Section~\ref{models} introduces our Metaverse AR socialization and information update system and UE-MSPCS allocation models. Then Section~\ref{RL} presents our proposed deep reinforcement learning approach. In Section~\ref{experiment}, extensive experiments are performed, and various methods are compared to show the satisfactory performance of our strategy. Section~\ref{conclude} concludes the paper.

\section{System model}
\label{models}

\begin{figure}[t]
\centering
\includegraphics[width=1\linewidth]{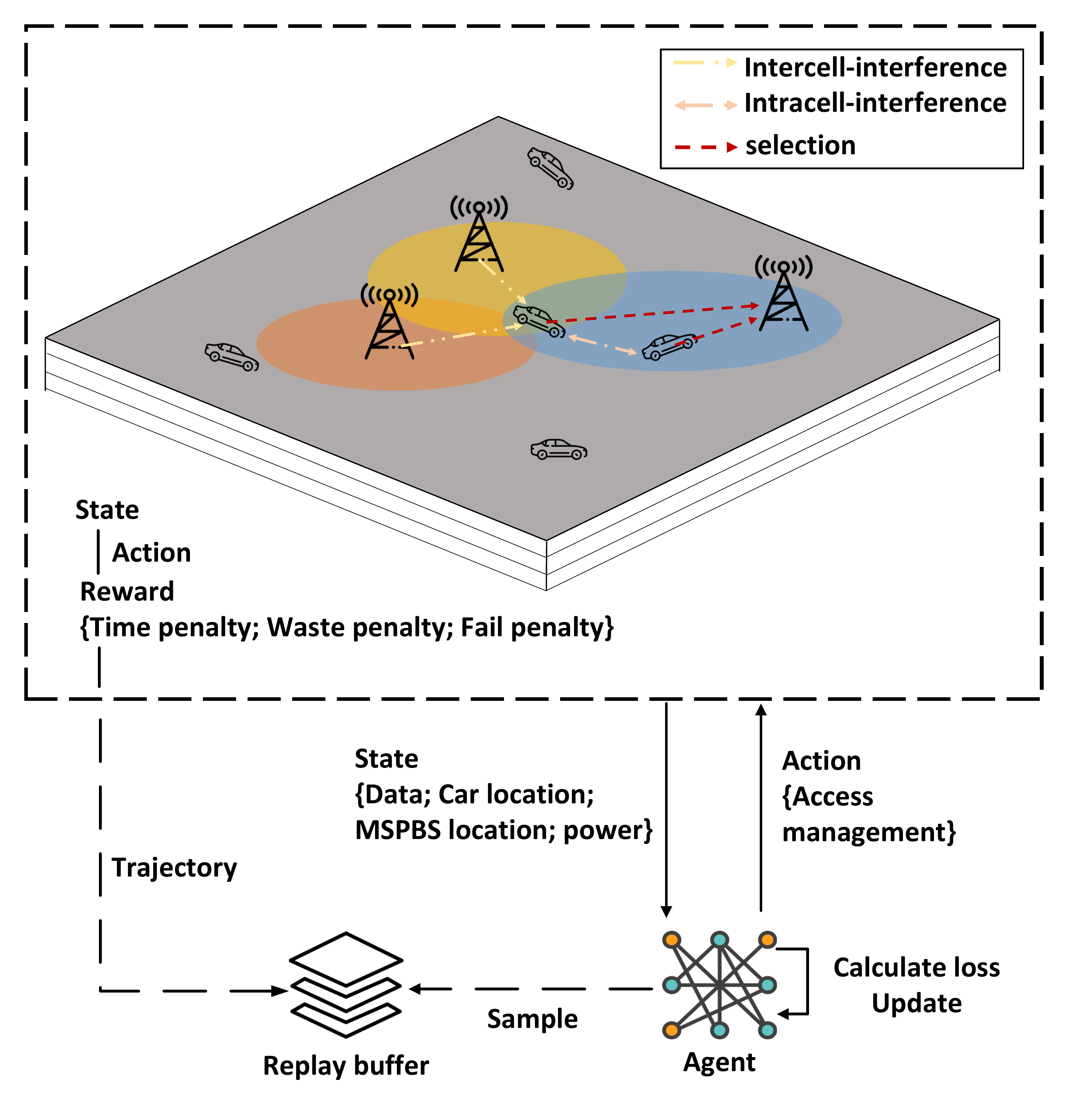}
\caption{System Model illustrating the interaction of agent and the environment introduced in our work.}
\label{fig:model}
\vspace{-0.4cm}
\end{figure}

Consider the downlink transmission of $\mathcal{N}=\{1,2,...,N\}$ user equipments (UEs) travelling in cars, consuming virtual world content in real-time on the go, from a set of $\mathcal{M} = \{1,2,...,M\}$ Metaverse Sevice Provider Cell Station (MSPCS). Each UE $i \in \mathcal{N}$ downloads virtual world scenes from an MSPCS. As the virtual world scenes are of large data sizes, the scenes are downloaded as UE moves and may not be completely downloaded in a single step of movement and there may be a handover of the download task from one cell station to another. Since there are several MSPCSs located geographically across our environment, we consider inter-cell and intra-cell interference, which may influence the data rates and consequently size of data transmitted. The UEs move randomly around the geographical space we defined, and their distance with respect to other UEs and MSPCSs also influences the data transfer rates. Each UE must be allocated to an MSPCS and uploads its scene to its assigned cell station in each round. Evidently, the choice of UE-MSPCS allocation influences data transfer rates and, consequently the time taken to complete the virtual scene data transmission. We next introduce the UE-MSPCS communication model to illustrate the scene described above. Our system model is illustrated in Fig.~\ref{fig:model}.

\subsection{Communication Model}\label{sub:Communication-Model}
Our communication model is based on a wireless cellular network. Each  MSPCS from $\mathcal{M} = \{1,2,...,M\}$  will manage its downlink channels with all UEs $\mathcal{N}=\{1,2,...,N\}$. Furthermore, we denote $D^{t} = \{D_{1}^{t},D_{2}^{t},...,D_{N}^{t}\}$ as the size of the remaining virtual world scene data to be transmitted from the MSPCS to the UEs at a discrete time step $t \in \{1,2,\ldots\}$. Specifically, $D_{i}^{t}$, $i \in \mathcal{N}$ is the  data needed to be transmitted from an MSPCS to UE $i$ at the beginning of time ${t}$. We denote $\textbf c^{t}=(c_1^t, ..., c_N^t)$ as the channel allocation tuple, where $c_i^t=v~(i \in \mathcal{N}, v \in \mathcal{M})$ (or an indicator variable $c^{t}_{i,v}$ taking $1$) denotes that UE $i$ is allocated to MSPCS $v$ at time $t$. Considering both the  intra-cell and inter-cell interference, we can derive the \textit{signal-to-interference-plus-noise ratio} (SINR) of UE $i$ at time $t$ as:
\begin{align}
    \Gamma_i^t=\frac{g_{v,i}p_{v,i} }{ g_{v,i}  \sum\limits_{n \in\mathcal{N} \setminus \{i\}:c_n^t=c_i^t}p_{v,n}  + \sum\limits_{j \in\mathcal{M} \setminus \{v\}} g_{j,i} \sum\limits_{k: c_k^t = j} p_{j,k}+w\sigma^{2}}.\label{eq:R1}
\end{align}
where $g_{v,i}$ is the channel gain between CS $v$ and UE $i$ (CS is short for MSPCS), $p_{v,i}$ is the transmit power of CS $v$ for communication with UE $i$, $p_{v,n}$ is the power allocated by CS $v$ for communicating with user $n \neq i$, $g_{j,i}$ is the channel gain between cell station $j \neq v$ and UE $i$,  $p_{j,k}$ is the power transmitted from cell station $j \neq v$ for communicating with other UE $k$, and $w$ is the bandwidth that each MSPCS can use, and $\sigma^2$ is the one-sided noise power spectral density.

Essentially, $g_{v,i}p_{v,i}$ is the signal strength of transmission by MSPCS $v$ at UE $i$, $\left [g_{v,i} \cdot \sum\limits_{n \in\mathcal{N} \setminus \{i\}:c_n^t=c_i^t}p_{v,n}\right]$ is the intra-cell interference caused by the signals of MSPCS $v$ for other UEs $n \neq i$, and $\left[\sum\limits_{j \in\mathcal{M} \setminus \{v\}} g_{j,i} \sum\limits_{k: c_k^t = j} p_{j,k}\right]$ is the inter-cell interference caused by the signals of all other MSPCSs $j \neq v$ for all other UEs $k$.

For each time step $t$ that there are still data remaining to be transmitted from the MSPCSs to the UEs, we consider that the data downlink process is not complete, and the iteration continues. Then the data left to be received by UE $i$ as:
\begin{align}
\setlength{\belowdisplayskip}{4pt plus 1pt minus 1.0pt} 
\setlength{\belowdisplayshortskip}{4pt plus 1pt minus 1.0pt}
\setlength{\abovedisplayskip}{4pt plus 1pt minus 1.0pt} \setlength{\abovedisplayshortskip}{0.0pt plus 2.0pt}
&D_{i}^{t+1}=D_{i}^{t}-r_{i}^{t}\cdot \Delta,
\end{align}
where
\begin{align}
&r_{i}^{t}=w\cdot \log_2 \left(1+\Gamma_i^t\right),\label{eq:R2}
\end{align}

The $\Delta$ represents the length of each time step. $r^t_i$ denotes the data transfer rate from an MSPCS to UE $i$ at time step $t$. From Equation (\ref{eq:R1}), it is evident that the presence of several transmitting cell stations causes interference to UEs, and the closer the distance a UE is to a base station that it is not connected to, the smaller the SINR and hence data transfer rate. Similarly, the connection of too many UEs to a single base station causes intra-cell interference, which also reduces the SINR and data transfer rate as well. Therefore, the quest for an optimal UE-MSPCS allocation management solution is of utmost importance. The notations used in this work are summarized in Table~\ref{table:notation} of the Appendix.

\subsection{Problem formulation}

Our goal is to find the optimal UE to MSPCS allocation arrangement within $T$ steps and to minimize the time taken for all UEs to download their requested virtual world scenes from the MSPCS.  We formulated our data transfer objective function as:

\begin{align}
\setlength{\belowdisplayskip}{4pt plus 1pt minus 1.0pt}
\setlength{\belowdisplayshortskip}{4pt plus 1pt minus 1.0pt}
\setlength{\abovedisplayskip}{4pt plus 1pt minus 1.0pt} \setlength{\abovedisplayshortskip}{0.0pt plus 2.0pt}
\min_{\boldsymbol{c^{t}}}&\max_{i \in \mathcal{N}}\{T_i\}, \label{eq:M1}\\
s.t.~~& D_i^1-\sum_{t=1}^{T_i}r_i^t\cdot \Delta \leq 0,  ~\forall i \in \mathcal{N},\\
& \sum_{j=1}^{M}c_{i,j}^{t}=1, ~\forall i \in \mathcal{N}, \forall t \in \mathcal{T}, \\
& \sum_{i=1}^{N}c_{i,j}^{t}>0, ~\forall j \in \mathcal{M}, \forall t \in \mathcal{T}, \\ &T_i \in \mathbb{N},  ~\forall i \in \mathcal{N}, \\
&c^{t}_{i,j} \in \{0,1\},  ~\forall i \in \mathcal{N},
\forall t \in \mathcal{T}.
\end{align}
where $T_i$ denotes the total time taken for UE $i$ to download their requested virtual world scenes from a cell station, $\mathcal{T}$ denotes the set of time steps, and $\mathbb{N}$ denotes the set of all natural numbers. We represent $T$ as $T:=\max_{i \in \mathcal{N}}\{T_i\}$. Intuitively, $T$ is the maximum time taken for all the UEs to download their  requested virtual world scenes from a cell station. Constraint (5) specifies that  UE $i$ has to complete the download of its requested virtual scene, with no data remaining to be transferred after time step $T_i$. Constraint (6) indicates that each UE can  be allocated to only one MSPCS, while constraint (7) indicates that each MSPCS needs to have at least 1 UE assigned to it. Constraints (8) and (9) constrain $T_i$ to fall within the set of natural numbers and $c^{t}_{i,j}$ to be binary, respectively.

Problem (\ref{eq:M1}) encompasses a time component, as packages are continuously and sequentially sent from the MSPCSs and UEs. Model-based methods cannot accommodate the time-sequential and randomly evolving nature of our proposed environment, and are therefore impractical for implementation.  Some similar works \cite{onetime} merely focus on optimizing one-time transmission arrangements due to the formidable and intractable complexity of time-sequential problems. Reinforcement learning is a suitable technique for solving sequential problems~\cite{RL1}. Hence, in the next section, we propose multiple feasible and effective deep reinforcement learning algorithms to solve our proposed problem. 

\section{Deep reinforcement learning approach}
\label{RL}
We have introduced our communication model and problem formulation in earlier sections. However, the proposed problem formulation has to be adapted and reformulated as a deep reinforcement learning problem. In this section, we will introduce our deep reinforcement learning approach, the design of states, actions, and structure of our algorithms.
\subsection{Reinforcement learning approach to our problem}\label{RLapproach}
An ingenious design of the state, action spaces and reward function is key to successfully adopting reinforcement learning methods in solving optimization problems. Now, we will expound on our state, action space, and reward design.
\subsubsection{State}
Although sophisticated states provide the agent with more information and a more comprehensive view of the environment, they can introduce complexity which may cause training to be erratic. Therefore, the number of features included in the state needs to be limited, and filtering out less relevant features is essential.


In our work, it is essential to include features which significantly influence UE-MSPCS allocation arrangement. This is especially so if the feature is evolving with time. Therefore, we include 1) virtual world scene data requested by UEs: $D_{i}^{t},~\forall i$, 2) channel gain between UEs and MSPCS: $g_{i,v}^{t},~\forall i,v$, and 3) power output by cell station allocated to UE: $p_{i},~\forall i$ into the state. These features greatly influence the transmission rates and consequently, UE-MSPCS allocation. Since the state dimensions increase drastically with an increasing number of UEs, we fix our state at these three attributes:
\begin{align}
    &S^{t} = \{D_{i}^{t},g_{i,v}^{t},p_{i}~|~i\in \mathcal{N},v \in \mathcal{M}\},
    ~t\in \mathcal{T}.\label{state}
\end{align}

\subsubsection{Action}
The action space is as delicate as the state and requires intricate design. It decides the available methods and available algorithms we can use. In our work, the reinforcement learning agent's action is to decide the UE-MSPCS allocation. We formally define the action space as:
\begin{align}
    &A^{t}=\textbf{c}^{t}= \{c^{t}_{i,v}~|~i \in \mathcal{N},v \in \mathcal{M}\} ,~ t\in \mathcal{T}.\label{action}
\end{align}
The number of the discrete actions is $N^{M}$, where $N$ denotes the total number of UEs and $M$ is the total number of MSPCS. The action space is tremendous when $N$ and $M$ are large.

Fortunately, large discrete and continuous action spaces are well studied, and there are several state-of-the-art algorithms which are able to tackle complex problems. We discuss these algorithms in greater detail in \ref{algorithms}.

\subsubsection{Reward}
Environments with sparse rewards typically impede training progress. Therefore, in our work, we design our rewards assignment in a way such that the agent receives more feedback through more consistent, intricately structured rewards. This aids exploration and training.

As the goal for our work is to minimize the total time taken for all UEs to download their requested virtual world scenes from the MSPCSs, we set the reward function as follows:
1) Time spent penalty: we give a reward of $-1$ for each time step where there is remaining virtual world scene to be downloaded from the MSPCSs; 2) Resource waste penalty: We give a reward of $-3$ for actions which cause UEs to download data when their requested remaining transferable data sizes are 0; 3) Fail penalty: we give a huge negative reward of $-100$ if the agent is unable to finish its task in 100 time steps. The inclusion of a fail penalty ensures that poor UE-MSPCS allocation and solution that may result in indefinitely long episodes do not hold up the overall training process.

\subsection{Reinforcement learning algorithms}
\subsubsection{\textbf{Basic structure}}
Q learning and Sarsa \cite{RLintro} are among the fundamental algorithms in dealing with the Markov decision process (MDP) problems. By observing the state $S$, an agent takes action $A$, causing changes in the environment and obtaining its reward $R$. This leads the agent to the next state $S'$. Sarsa is on-policy which needs to get $A'$ before updating the network, whereas Q learning always chooses the estimated optimal action as the $A'$. Thus, the equation update of Sarsa and Q-learning are expressed as such:
\begin{align}
    &Q(S^{t},A^{t}) \leftarrow Q(S^{t},A^{t})+\alpha[target-Q(S^{t},A^{t})].
\end{align}
where
\begin{align}
    &target = R^{t+1}+\gamma Q(S^{t+1},A^{t+1})~~~~~\text{(Sarsa)},\\
    &target = R^{t+1}+\gamma \max_{A^t}Q(S^{t+1},A^t)~~~~\text{(Q-learning)}.
\end{align}
Here, $Q(S^{t},A^{t})$ denotes the estimated value of taking action $A$ upon arriving at state $S$ at time $t$, and $\gamma$ is the discount factor. 
\subsubsection{\textbf{Actor-Critic}}
The actor-critic \cite{mnih2016asynchronous} reinforcement learning structure is considered as among the state-of-the-art frameworks, and is commonly used to tackle wireless communication problems \cite{shah2021joint}. It utilizes an actor based on policy gradient (PG) to choose an action and a critic network for estimating the Q-values. The actor-critic structure forms the underlying structure of many state-of-art algorithms, such as DDPG\cite{lillicrap2015continuous}, SAC\cite{haarnoja2018soft}. The Actor has its policy $\pi$ of selecting actions, typically represented by a convolutional neural network (CNN) with parameter $\theta$. However, as our state design is not as sophisticated, we utilize a Fully Connected Neural Network (FCNN) parameterized by $\theta$ for the Actor. In policy gradient algorithm, the agent samples trajectories $\tau=\{S^{1},A^{1},S^{2},A^{2},...\}$. The $\tau$ is not deterministic as the agent may select different actions under a specific state, so, if we assume the probability of a certain $\tau$ is $p_{\theta}(\tau)$, we have the expected reward as:
\begin{align}
    \Bar{R}_{\theta}=\sum_{\tau}R(\tau)p_{\theta}(\tau). \label{drleq1}
\end{align}
In the actor-critic network, instead of $R$, we use Q-value and create a critic network for estimating it. According to (\ref{drleq1}), the policy gradient update in Actor is:
\begin{align}
    \nabla \Bar{R}_{\theta}=\frac{1}{N} \sum_{i-1}^{N} \sum_{t=1}^{T_{i}}Q^{\pi_{\theta}}(S^{t}_{i},A^{t}_{i})\nabla \log p_{\theta}(A^{t}_{i}|S^{t}_{i}).
\end{align}
Simultaneously, the critic network updates through the Mean Square Error (MSE) between estimated Q-value and factual Q-value. The loss is defined as:
\begin{align}
    loss=\frac{1}{N}\sum_{i-1}^{N} \sum_{t=1}^{T_{i}}(R^{t}_{i}+\max_{A^{t+1}_{i}}Q^{\pi_{\theta}}(S^{t+1}_{i},A^{t+1}_{i})-Q^{\pi_{\theta}}(S^{t}_{i},A^{t}_{i}))^{2}.
\end{align}
With this structure, the actor-critic network is capable of handling complex environments with continuous state and action spaces.
\subsubsection{\textbf{Advantage Actor Critic}}
Advantage actor-critic structure is a derivative of actor-critic. Similar to the action-value $Q^{\pi}(S,A)=\mathbb{E}[R^{t}|S^{t}=S,A]$ of the actor-critic framework, which is the expected return for selecting $A$
 at $S$ following policy $\pi$, we can correspondingly use $V^{\pi}(S)=\mathbb{E}[R^{t}|S^{t}=S]$ to denote the value of state $S$. Therefore, in the advantage actor-critic architecture, it is common to use $Adv(A^{t},S^{t})=Q(A^{t},S^{t})-V(S^{t})$ as the $advantage$ of action $A^{t}$ under state $S^{t}$.
 
\subsection{Algorithms} \label{algorithms}
We utilize four commonly used deep reinforcement learning algorithms; DQN, dueling DQN, A2C, and PPO. We explain them as follows.

\subsubsection{\textbf{Deep Q network (DQN)}} Our proposed scenario requires a massive action space which renders traditional Q-learning and Sarsa impractical. Therefore, for our work, we utilized DQN \cite{RL1} based on Q-learning which uses a deep neural network to handle a continuous state space and massive action space. DQN relies on a replay buffer to store its trajectories and samples these trajectories for training. An improved version of DQN is the renowned dueling DQN (DDQN) \cite{duelDQN} which utilizes two heads to compute a scalar state value ($V$) and the advantages ($A$) for each action. Both the $V$ and $A$ are then utilized to compute the Q-value. In this work, we utilized both DQN and DDQN to solve our proposed problem.

\subsubsection{\textbf{A2C}}

Advantage actor-critic (A2C) \cite{A2C} utilizes multiple workers and processes to sample data to train their own policies within their own environments. Each worker of the A2C framework will then synchronously upload their newly updated parameters to the global network. The global network will then update them with a unified parameter. One significant advantage of A2C is that it can take advantage of multi-cores on computers for computation, which reduces training time dramatically. Furthermore, with the advantage actor-critic framework, lower training variance and improved robustness can be achieved. Another advantage of A2C is that A2C can handle very complex environments and scenarios and hence is adopted for our work.



\subsubsection{\textbf{Proximal Policy Optimization (PPO)}}
Proximal Policy Optimization (PPO)~\cite{schulman2017proximal} is an advanced reinforcement learning method proposed by openAI, and it is an improvement over traditional policy gradient methods. PPO can handle sequential problems faced in reinforcement learning very well, as it adopts a conservative approach of utilizing Kullback Leibler (KL) divergence to constrain the magnitude of policy change.

\section{Experiments}
\label{experiment}
In this section, we will first discuss our environment settings. We then provide details of the hyper-parameters used in our experiments. Finally, we provide in-depth analysis of our results.

\subsection{Configuration} \label{Configuration}
We configured three network settings: (i) 4 UEs and 3 MSPCSs, (ii) 6 UEs and 3 MSPCSs, and (iii) 7 UEs and 4 MSPCSs to test our proposed approaches and algorithms. The bandwidth and noise are simulated to be $w=10$ MHz and $\sigma^2=-100$ dBm. We conduct experiments in $10^4$ episodes, where each episode has an entire execution of the optimization problem~(\ref{eq:M1}) over multiple time steps.  We then set different power output at each time step for different UEs, which falls between $(0.5,2.0)$ Watt. Each UE's virtual environment size ($D_i^1$) varies every episode from $100$ to $300$ Mb. Thus we set the values to be randomly generated from the given data size range. Each time step represents 5 seconds. The locations of UEs and MSPCSs are set as follows. At the beginning of each episode, each UE and each MSPCS are placed uniformly at random in the simulation environment, which is a $1000\text{m} \times 1000\text{m}$ area. Then the mobility model of each UE across different time steps of an episode is as follows: at the end of each time step, UEs randomly move a maximum of $100$m in $x$ and $y$ directions, in which $x$ and $y$ represent the longitudinal and latitudinal directions in the $1000\text{m} \times 1000\text{m}$ map. For simplicity, we use the free space path loss model \cite{doble1996introduction} for the channel gain between UE $i$ and MSPCS $v$ at time step $t$:
\begin{align}
g_{i,v}^{t}=\left(\frac{\lambda}{4\cdot \pi\cdot dist_{i,v}^{t}}\right)^2.\label{eq:R5} 
\end{align}
where $dist_{i,v}^{t}$ denotes the distance between UE $i$ and MSPBS $v$, while $\lambda$ represents the wavelength of the transmission at time step $t$. We consider terahertz (THz) for 6G communications, so we set $\lambda$ as 0.3mm (the wavelength for 1THz). 

\subsection{Implementation}
We implement four algorithms introduced in Section~\ref{algorithms}. Note that for every algorithm with a replay buffer, we first sample data with a random policy to warm up (fill replay buffer with data), which is a ubiquitous but effective initialization strategy. In addition, we also use a random policy that chooses actions randomly each time for comparison. Simultaneously, the Adam optimizer\cite{adam} is adopted for all our implemented algorithms. To better observe the final performance, we train the models with 40000 episodes for network setting (i), setting (ii) and setting (iii) (the three settings have been explained in Section~\ref{Configuration} above).

After multiple trials and extensive parameter adjustments, we finalize and list the critical hyper-parameter settings for each algorithm in Table \ref{table:parameter} of the Appendix.

\subsection{Result analysis}

\begin{figure}[t]
\centering
\subfigtopskip=0pt
\subfigbottomskip=0pt
\subfigure[3 MSPCSs and 4 UEs.]{
\begin{minipage}[t]{1\linewidth}
\centering
\includegraphics[width=.95\linewidth]{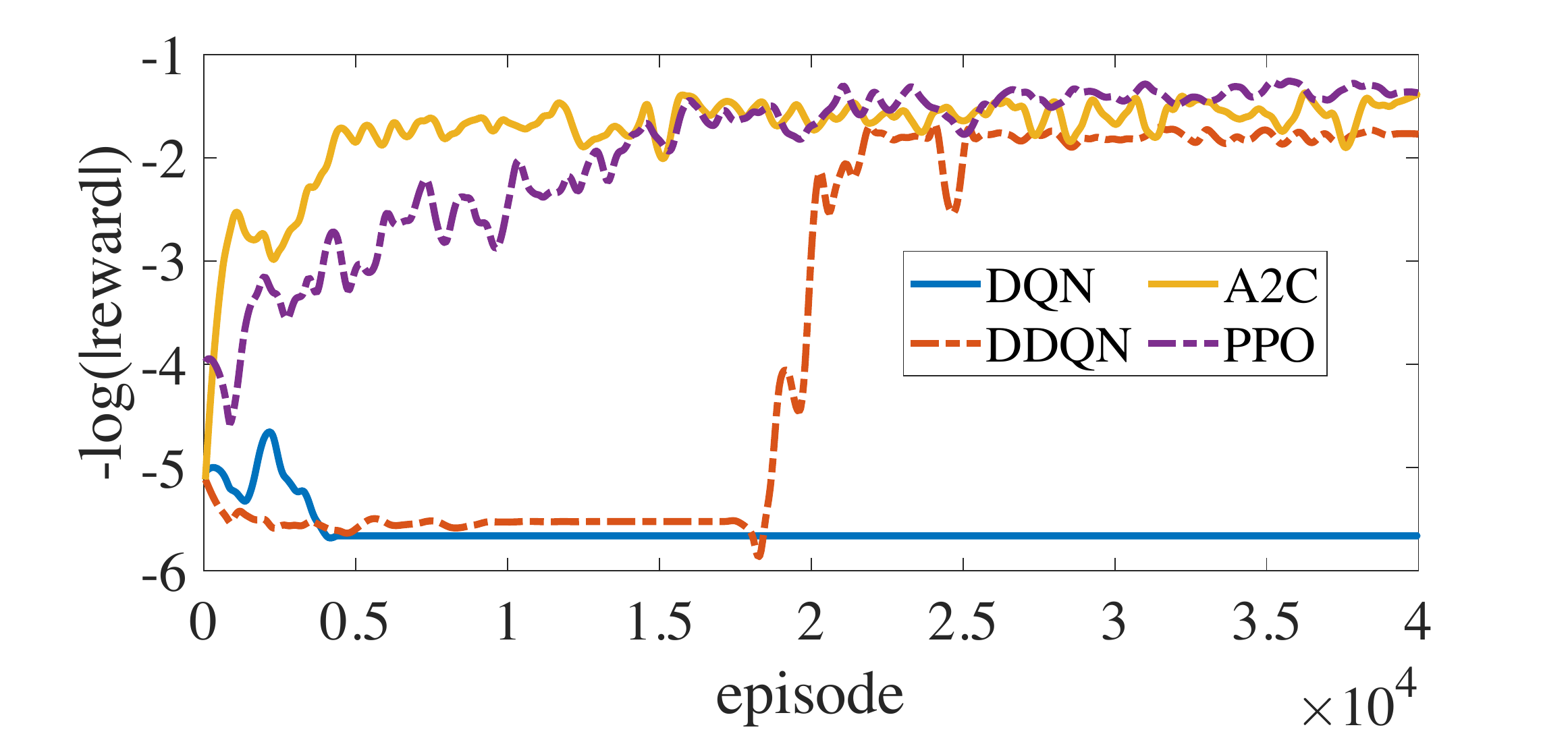}
\label{fig:43}
\vspace{-20pt}
\end{minipage}%
}%

\subfigure[3 MSPCSs and 6 UEs.]{
\begin{minipage}[t]{1\linewidth}
\centering
\includegraphics[width=.95\linewidth]{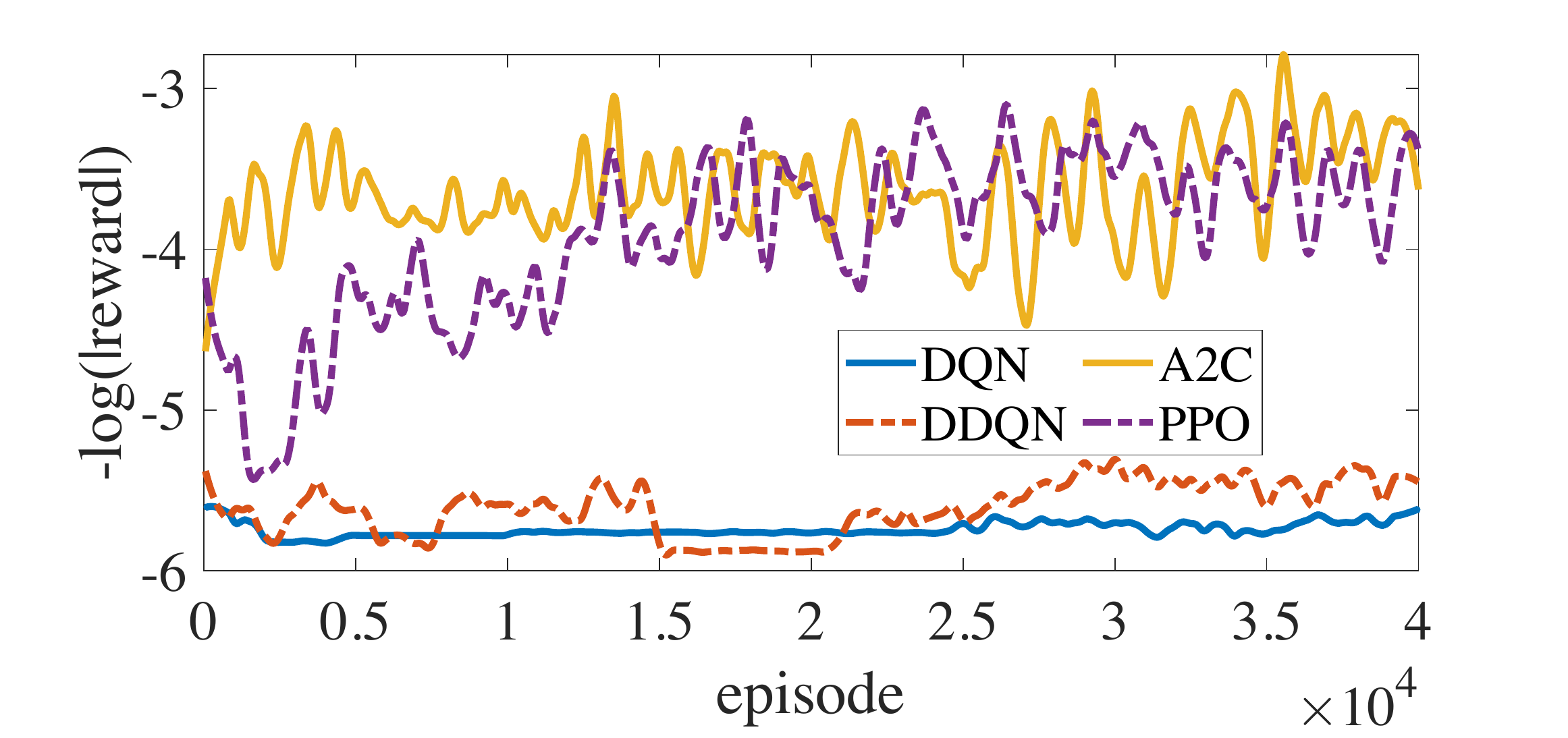}
\label{fig:63}
\vspace{-20pt}
\end{minipage}%
}%

\subfigure[4 MSPCSs and 7 UEs.]{
\begin{minipage}[t]{1\linewidth}
\centering
\includegraphics[width=.95\linewidth]{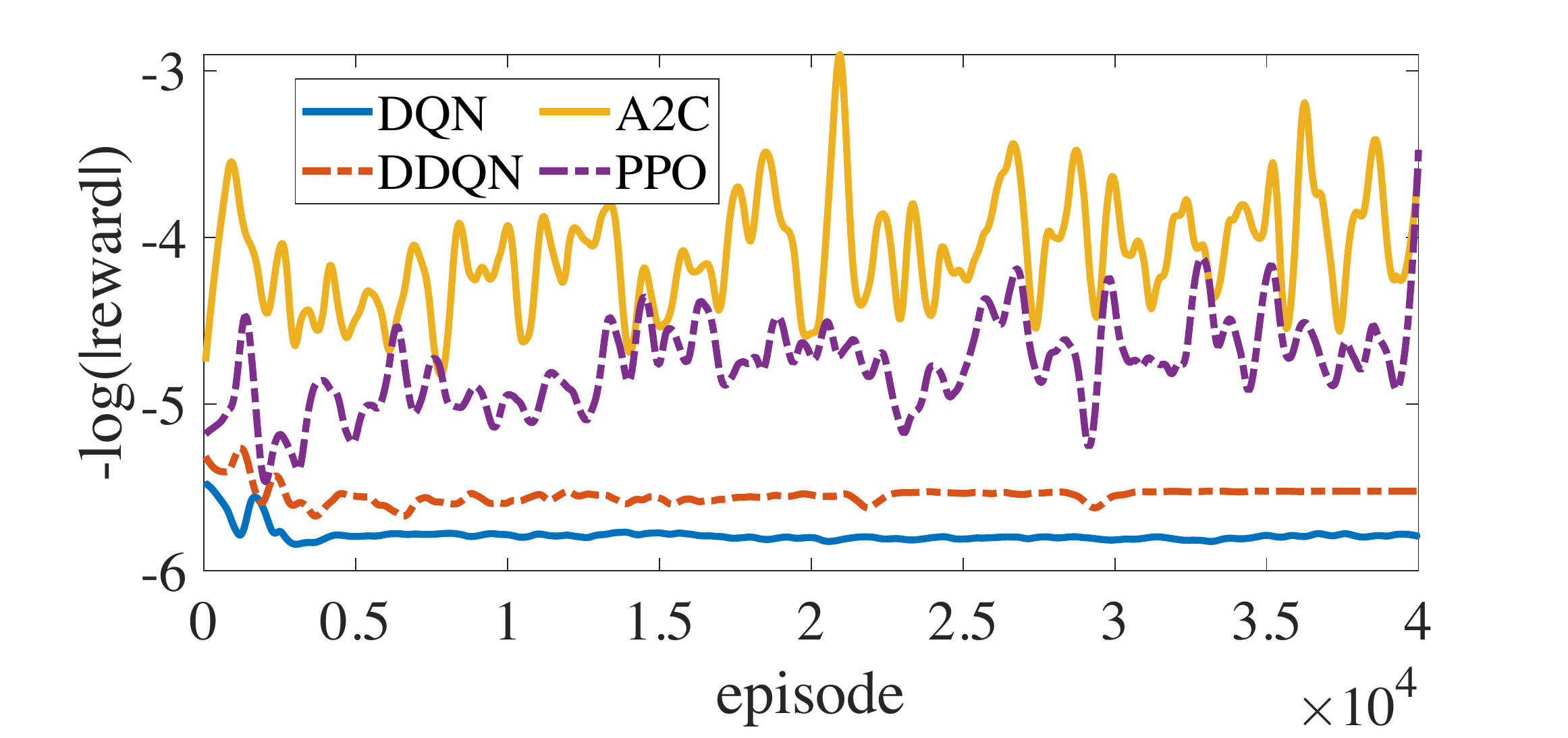}
\label{fig:83}
\vspace{-20pt}
\end{minipage}
}%
\caption{Various algorithms' $-\log(|\text{reward}|)$ under different network settings.} 
\vspace{-10pt}
\end{figure}

Four algorithms (DQN, DDQN, A2C, PPO) under three different scenarios are tested to compare our proposed method. 

\begin{figure}[t]
\centering
\subfigtopskip=0pt
\subfigbottomskip=0pt

\subfigure[Steps taken in each episode during training.]{
\begin{minipage}[t]{0.9\linewidth}
\centerline{\includegraphics[width=1\linewidth]{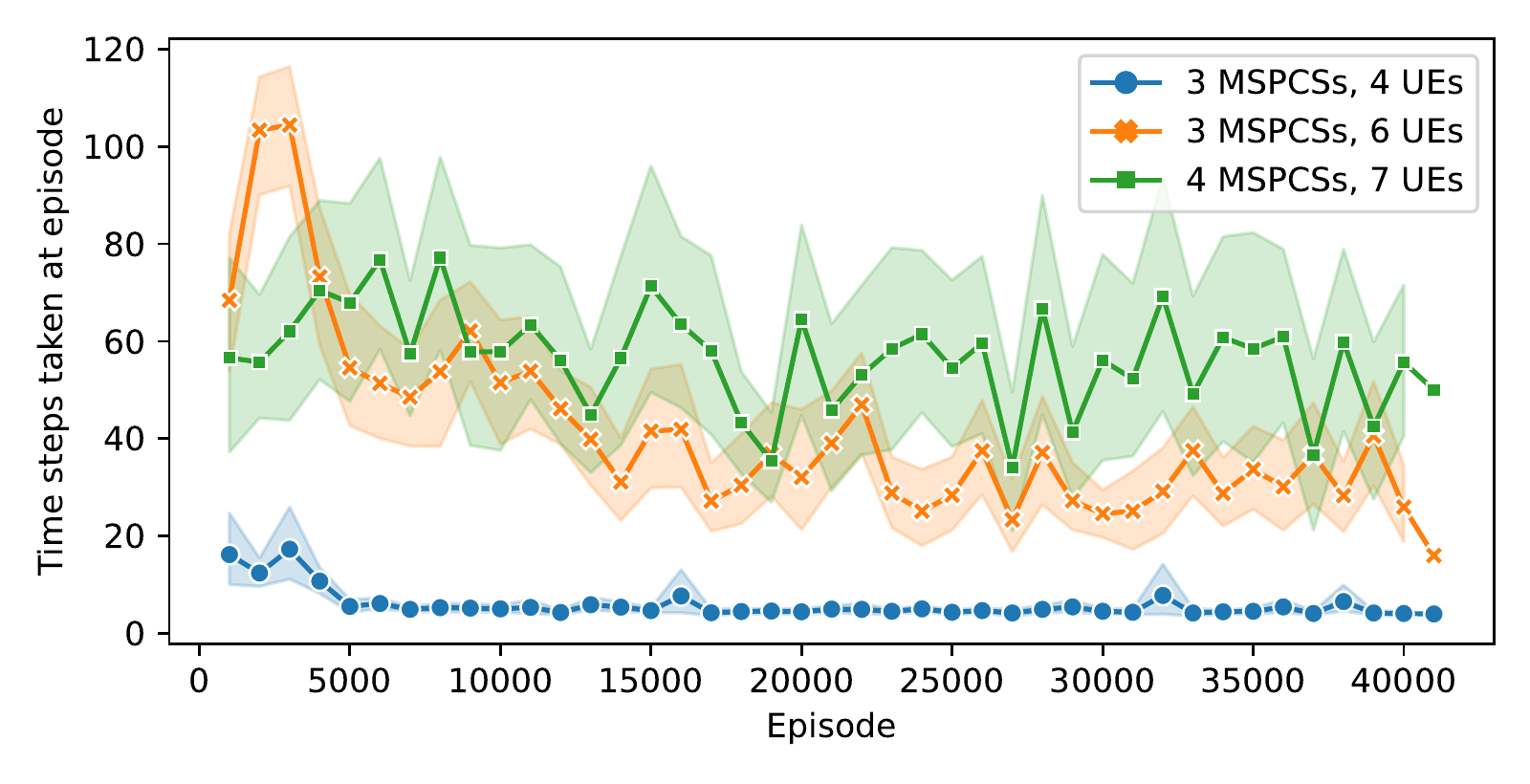}}
\label{fig:steps}
\end{minipage}
}

\subfigure[The resource waste counts of all UEs during training.]{
\begin{minipage}[t]{0.9\linewidth}
\centerline{\includegraphics[width=1\linewidth]{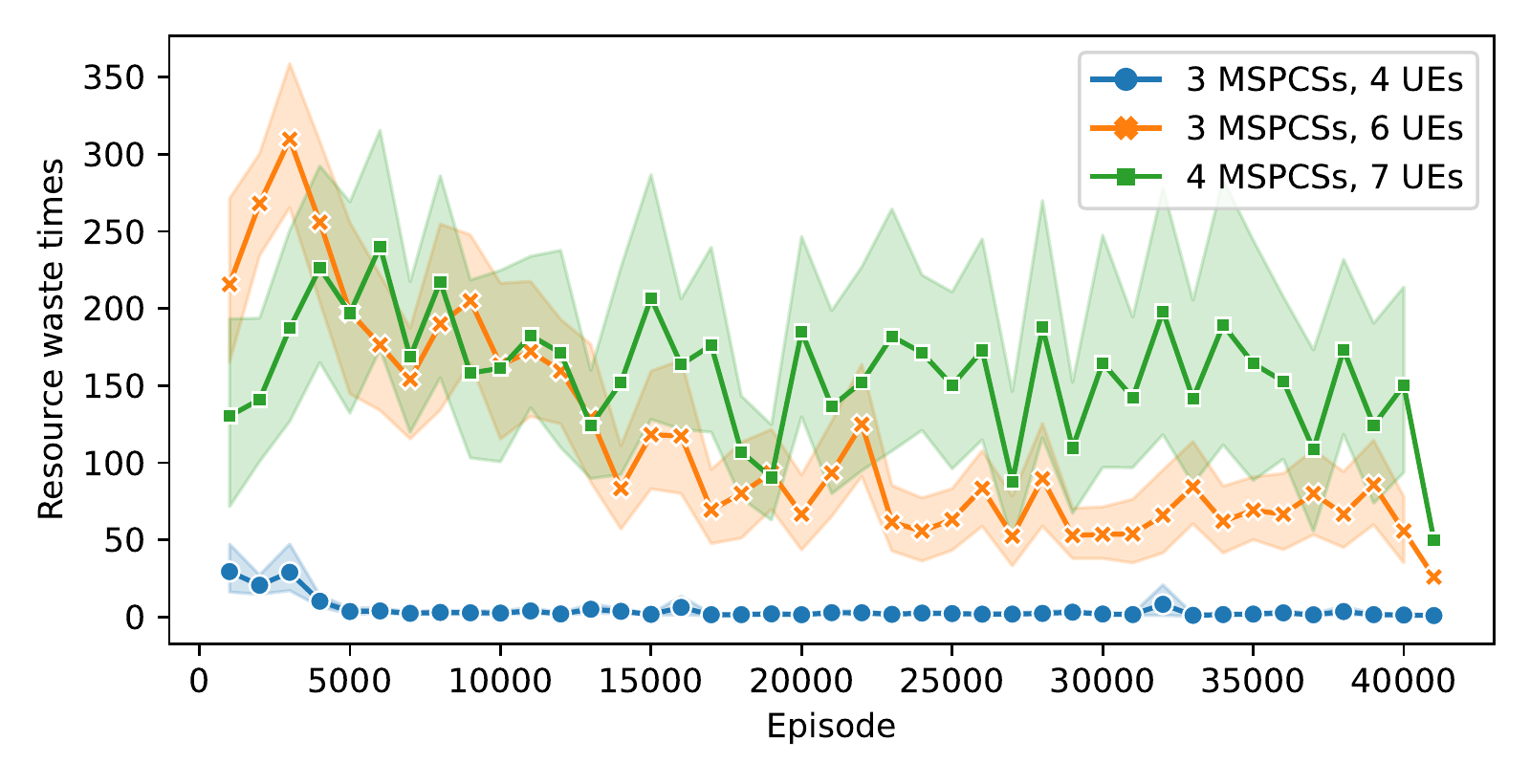}}
\label{fig:waste}
\end{minipage}
}
\caption{Steps of episodes and resource waste counts.}
\vspace{-10pt}
\end{figure}


As shown in Fig.~\ref{fig:43} to Fig.~\ref{fig:83}, A2C performs the best among them. It always attains the highest eventual reward, which can be attributed to its advantage actor-critic base structure and superior sample efficiency, as it uses multiple agents for sampling. PPO comes in as a runner-up. Although PPO is considered a more stable algorithm due to its policy-constrain feature, it fails to find a near-optimal solution in our proposed scenario under the  network setting with the highest UE count. DQN performs the worst of the algorithms due to its inability to handle complex environments when the state and action space are enormous. Finally, we observe that DDQN performs much better than vanilla DQN, as we can see that it obtained a satisfactory reward in the end under the  network setting with the smallest UE count.

As the A2C algorithm performs the best in each network setting, we further study other metrics under the A2C framework. Fig.~\ref{fig:steps} and Fig.~\ref{fig:waste} illustrate the required number of steps to deliver all of the UEs' requested data and the resource wastage counts, respectively, with the A2C algorithm. Each resource wastage count is defined as a UE being allocated to an MSPCS despite having no requested data remaining. We performed the experiments at multiple seeds (10 different seeds) and recorded these metrics every 100 episodes. The faint outer bands represent the confidence interval. We can see that as the training steps increase, the maximum $T$ and resource wastage counts both declines, indicating that our deep reinforcement learning approach is improving and converging to a near-optimal solution.

\section{Conclusion}
\label{conclude}
 We present a novel Metaverse socialization model which aims to model mobile users travelling in vehicles on-the-go and downloading virtual world scenes from the MSPCSs in real-time. Traditional optimization strategies are not suitable to tackle this problem due to the time-sequential nature of the problem. Hence, we proposed deep reinforcement learning approaches in this work. Multiple algorithms are compared, and the experiments demonstrate that A2C performs best in our proposed scenario. PPO struggled to find a near-optimal solution in a more complicated scenario, but it is more stable due to its policy-constrain feature.
 Our study can be viewed as a use case of the Metaverse over AI-enabled 6G communications. A future direction is to investigate other ways where AI can be used to improve 6G communications for the Metaverse.

\section*{Acknowledgement}

This research is supported in part by Nanyang Technological University Startup Grant; in part by the Singapore Ministry of Education Academic Research Fund under Grant Tier 1 RG97/20, Grant Tier 1 RG24/20 and Grant Tier 2 MOE2019-T2-1-176.

{\small



}

\appendix

~\vspace{-20pt}


\begin{table}[h]
\caption{Notations\vspace{-10pt}}
\label{table:notation}
\begin{center}
\begin{tabular}{|l|p{7cm}|}
\hline
{}{Symbol}&{}{Description} \\\hline
{}{$n,m,t$}&{}{Index of UEs, MSPCS and time step}\\
{}{$\mathcal{M}$}&{}{Set of MSPCSs}\\
{}{$\mathcal{N}$}&{}{Set of UEs}\\
{}{$D^t$}&{}{Remaining data to transmit to each UE at time step $t$}\\
{}{$\mathcal{T}$}&{}{Total time steps used to send all requested virtual world  data}\\
{}{\textbf{$c^t$}}&{}{Access management at time $t$}\\
{}{$\Delta$}&{}{Duration of one time step.}\\
{}{$r_n^t$}&{}{Transmission rate of UE $n$ at time $t$}\\
{}{$w$}&{}{Bandwidth}\\
{}{$p_v,i$}&{}{Power of MSPCS $v$ used to transmit to UE $i$}\\
{}{$g_{i,v}^t$}&{}{channel gain between UE $i$ and MSPCS $v$ at time $t$}\\
{}{$\sigma$}&{}{noise parameter}\\
{}{$\Gamma_i^t$}&{}{SINR of UE $i$ at time $t$}\\
\hline
\end{tabular}
\vspace{-25pt}
\end{center}
\end{table}

\begin{table}[h]
\centering
\caption{Important hyper parameters}
\label{table:parameter}
\vspace{-5pt}
\scalebox{0.8}{
\begin{tabular}{ccccccc}
\hline
 \begin{tabular}[c]{@{}l@{}}network\\ setting\end{tabular}  & \begin{tabular}[c]{@{}l@{}}learning\\ rate\end{tabular}& \begin{tabular}[c]{@{}l@{}}hidden\\ layers\end{tabular} & batch size & \begin{tabular}[c]{@{}l@{}}entropy\\ coefficient\end{tabular} & \begin{tabular}[c]{@{}l@{}}Generalized \\ Advantage\\ Estimator\\ (GAE)\end{tabular} & \begin{tabular}[c]{@{}l@{}}discount\\ factor\end{tabular} \\ \hline
  \multicolumn{7}{c}{Deep Q learning} \\ \hline
$4$ UEs, $3$ MSPCSs   & $1\times10^{-3}$ & $128$ & $64$ & & & $0.99$\\
$7$ UEs, $3$ MSPCSs   & $5\times10^{-4}$ & $128$ & $64$ & & & $0.99$\\
$7$ UEs, $4$ MSPCSs   & $5\times10^{-4}$ & $256$ & $64$ & & & $0.99$\\ \hline

  \multicolumn{7}{c}{DDQN} \\ \hline
$4$ UEs, $3$ MSPCSs   & $1\times10^{-3}$ & $128+64$ & $64$ & & & $0.99$\\
$6$ UEs, $3$ MSPCSs   & $5\times10^{-4}$ & $256+64$ & $64$ & & & $0.99$\\
$7$ UEs, $4$ MSPCSs   & $1\times10^{-4}$ & $256+64$ & $64$ & & & $0.99$\\ \hline

  \multicolumn{7}{c}{A2C} \\ \hline
$4$ UEs, $3$ MSPCSs   & $5\times10^{-4}$ & $256$ & $64$ & $1\times10^{-3}$ & $0.95$ & $0.99$\\
$6$ UEs, $3$ MSPCSs   & $5\times10^{-5}$ & $512$ & $64$ & $1\times10^{-3}$ & $0.95$ & $0.97$\\
$7$ UEs, $4$ MSPCSs   & $1\times10^{-5}$ & $512$ & $64$ & $1\times10^{-3}$ & $0.95$ & $0.97$\\ \hline

  \multicolumn{7}{c}{PPO} \\ \hline
$4$ UEs, $3$ MSPCSs   & $5\times10^{-4}$ & $128$ & $64$ & $1\times10^{-4}$ & $0.95$ & $0.99$\\
$6$ UEs, $3$ MSPCSs   & $1\times10^{-4}$ & $256$ & $64$ & $1\times10^{-4}$ & $0.95$ & $0.99$\\
$7$ UEs, $4$ MSPCSs   & $5\times10^{-5}$ & $512$ & $64$ & $1\times10^{-4}$ & $0.93$ & $0.95$\\ \hline
\end{tabular}
}
\vspace{-0.3cm}
\end{table}

\end{document}